\documentclass[prb,twocolumn,showpacs,amssymb,superscriptaddress]{revtex4-2}
\usepackage{graphicx}
\usepackage{amsmath}
\usepackage{amsfonts}
\usepackage{amsthm}
\usepackage{amssymb}
\usepackage{amsbsy}
\usepackage{wasysym}
\usepackage{bm}
\usepackage{bbm}
\usepackage{mathrsfs}
\usepackage{color}
\usepackage{hyperref}
\usepackage{braket}
\usepackage{soul}
\setcitestyle{super}

\date{\today}
\begin{document}

\title{Hydrodynamics of quantum spin liquids}

\author{Vir B. Bulchandani}
\affiliation{Princeton Center for Theoretical Science,  Princeton University, Princeton, New Jersey 08544, USA}

\author{Benjamin Hsu}
\affiliation{Amazon Web Services, Inc., Seattle, WA 98109}
\altaffiliation{Work done while the author was at Princeton University.}

\author{Christopher P. Herzog}
\affiliation{Mathematics Department, King’s College London,
The Strand, London, WC2R 2LS, UK}

\author{S. L. Sondhi}
\affiliation{Department of Physics, Princeton University, Princeton, New Jersey 08544, USA}

\begin{abstract}
Quantum spin liquids are topological states of matter that arise in frustrated quantum magnets at low temperatures. At low energies, such states exhibit emergent gauge fields and fractionalized quasiparticles and can also possess enhanced global symmetries compared to their parent microscopic Hamiltonians. We study the consequences of this emergent gauge and symmetry structure for the hydrodynamics of quantum spin liquids. Specifically, we analyze two cases, the $U(1)$ spin liquid with a Fermi surface and the $SU(4)$-symmetric ``algebraic'' spin liquid. We show that the emergent degrees of freedom in the spin liquid phase lead to a variety of additional hydrodynamic modes compared to the high-temperature paramagnetic phase. We identify a hydrodynamic regime for the internal $U(1)$ gauge field common to both states, characterized by slow diffusion of the internal transverse photon.
\end{abstract}

\maketitle

\section{Introduction}
Hydrodynamics provides a universal description of the long-wavelength, low-frequency dynamics of physical systems as they approach a state of thermal equilibrium. This dynamics is typically determined by a handful of classical partial differential equations, whose structure is dictated by the symmetries of the system at hand. The physical insight underlying this enormous reduction in complexity is the notion of a ``conservation law'': only local physical quantities that are conserved in every single microscopic collision can give rise to cumulative effects that are discernible on macroscopic scales. Hydrodynamics in this sense is equally applicable to both quantum and classical systems. Although the emergence of hydrodynamics in many-body classical systems has been appreciated for centuries, its emergence in many-body quantum systems is less well understood. Some notable exceptions include the theory of transport in metals and liquid helium-3, as based on Landau's Fermi liquid picture~\cite{lifshitz2013statistical,Shankar}, and the theory of superfluidity in weakly interacting Bose gases~\cite{GPELowDim,GPE}.

The idea that hydrodynamics might be a useful tool for understanding dynamics in a much broader range of quantum systems has only come into focus comparatively recently. Simultaneous advances in various experimental and theoretical directions have provided an intense stimulus for the study of hydrodynamics in many-body quantum systems. One active field of inquiry focuses on high-mobility two-dimensional electron gases, for which a long-anticipated regime of viscous electron flow~\cite{gurzhi1963minimum,GurzhiPotassium,MOLENKAMP1994551} has recently been realized in materials such as graphene~\cite{Bandurin1055} and $\mathrm{PdCoO_2}$~\cite{Moll1061}. Another rapidly growing field is the hydrodynamics of one-dimensional quantum systems, for which analytical tools based on integrability~\cite{CastroAlvaredo,Fagotti,Solvable} and numerical tools based on tensor-network methods~\cite{Znidaric,KPZ,DupontMoore} yield quantitative predictions in remarkable agreement with experimental results~\cite{Dubail,malvania2020generalized,Scheie2021}. These studies complement new theoretical methods for deriving hydrodynamics from the AdS-CFT correspondence~\cite{ADSCFT,KSS}, random quantum circuits~\cite{SpreadCK,SpreadN,SpreadVK} and effective field theory~\cite{Crossley2017,glorioso}.

In this paper, we explore the linear-response hydrodynamics of quantum spin liquids. These unusual topological states of matter arise in frustrated quantum magnets at low temperatures and exhibit non-trivial low-energy degrees of freedom~\cite{Savary_2016,SLRMP}. These degrees of freedom include fractionalized quasiparticles with enlarged flavour symmetries, and internal gauge fields coupled to these quasiparticle degrees of freedom. This enhanced low-energy structure can give rise to an elaborate structure of hydrodynamic slow modes compared to the high-temperature paramagnetic phase---even though neither phase breaks any symmetries of the microscopic Hamiltonian. The extra modes come in part from the emergent gauge fields and thus an emergent magnetohydrodynamics and in part from the new global flavour symmetries. 

We illustrate this behaviour by focusing on two representative examples of quantum spin liquid states, namely the $U(1)$ spin liquid with a Fermi surface, that is conjectured to describe the uniform resonating valence bond (RVB) state of the square-lattice Heisenberg antiferromagnet~\cite{SLRMP}, and the $SU(4)$-symmetric ``algebraic'' spin liquid, that describes its ``d-wave'' RVB state~\cite{HermeleStab}. While these are much studied spin liquids and hence appropriate objects of further investigation, our focus on these two dimensional examples compels us to address an additional subtlety arising from microscopic compactness of the emergent gauge field. This is the fact that $U(1)$ spin liquid states in two dimensions and at non-zero temperature are \emph{unstable} at the longest length and time scales, since the possibility of topological excitations of the internal gauge field ultimately leads to spinon confinement~\cite{HermeleStab}. (More generally, spin liquid states in $2+1$ dimensions with continuous gauge symmetry and sufficiently few fermion flavours are expected to exhibit confinement\cite{AppelQED,AppelQCD}.) For this reason it would be simpler to consider the hydrodynamic regime for $U(1)$ spin liquids in three dimensions, whose topological phases can persist to non-zero temperatures.

Thus the question arises of whether $U(1)$ spin liquids in two dimensions exhibit a hydrodynamic regime at all. We argue below that such a regime indeed exists, provided that monopole operators in $2+1$ dimensions are sufficiently irrelevant, as has been conjectured for both the $U(1)$ spin liquid with a Fermi surface~\cite{SSLee} and Dirac spin liquid states with many fermion flavours~\cite{ASL}. We find that despite their fundamental differences, the internal $U(1)$ gauge symmetry common to both these states gives rise to a diffusive hydrodynamic mode of the internal transverse photon, that is a spin-liquid analogue of magnetic diffusion in ordinary conductors~\cite{jackson1999classical,BAGGIOLI20201}. For chiral spin liquid states\cite{WWZ,KalmeyerLaughlin,Baskaran}, or more generally in the presence of a Chern-Simons term for the emergent gauge field, this hydrodynamic mode is suppressed (see Appendix \ref{sec:AppB}).

\section{Generalities}

\subsection{Models}

For concreteness, consider the family of antiferromagnetic, spin-$1/2$ Heisenberg models on a square, two-dimensional lattice,
\begin{equation}
\label{eq:MicHamiltonian}
H = J\sum_{\langle i j \rangle} \mathbf{S}_i \cdot \mathbf{S}_j + \ldots,
\end{equation}
where the ellipsis denotes terms consistent with both lattice symmetry and $SU(2)$ symmetry. At high temperatures, such models will exhibit standard paramagnetic behaviour. At low temperatures, they can potentially host a variety of spin liquid phases that depend on the microscopic details of the model in question. The field theory of these states can be studied with the aid of slave-fermion variables~\cite{Abrikosov}
\begin{equation}
S^j_i = \frac{1}{2} \sum_{\alpha,\beta}f_{i\alpha}^{\dagger} \sigma^j_{\alpha\beta} f_{i\beta},
\end{equation}
which yield an exact rewriting subject to the single-occupancy constraint $\sum_\alpha f_{i\alpha}^{\dagger}f_{i\alpha} = 1$. Various ``spin liquid" saddle points may then be studied at the level of mean-field theory. Physics in the vicinity of these saddle points is characterized by the emergence of fractionalized excitations and dynamical gauge fields associated with these excitations. One question that has not been studied systematically is how far these additional excitations and enhanced symmetries lead to novel signatures in the hydrodynamics of quantum spin liquids, compared to the high-temperature paramagnetic regime.

This is the question we address below, focusing on two representative examples of spin liquid states, the $U(1)$ spin liquid with a Fermi surface, corresponding to the uniform RVB , or ``zero flux state'', of the Heisenberg model\cite{SLRMP}, and the $SU(4)$-symmetric algebraic spin liquid, corresponding to the d-wave RVB, or ``staggered flux state''~\cite{ASL}. For each of these states, there is an effective low-energy field theory description consisting of multiple spinon degrees of freedom minimally coupled to a dynamical, non-compact, $U(1)$ gauge field. However, the physics of these two effective theories is very different. The field theory describing the spin liquid with a Fermi surface is neither Galilean nor Lorentz invariant, and there are no internal symmetries beyond the emergent $U(1)$ symmetry. Meanwhile, the field theory describing the algebraic spin liquid is Lorentz invariant, quantum critical, and exhibits multiple quasiparticle flavours with an internal $SU(4)$ symmetry. 

Despite these differences, both spin liquid states exhibit various hydrodynamic features in common, most notably a diffusive transverse mode of the internal gauge photon. In physical terms, this corresponds to an excitation of the internal magnetic field $\mathbf{b}$, and therefore couples to the ``spin chirality'' $\mathcal{P} = \mathbf{S}_1 \cdot \mathbf{S}_2 \times \mathbf{S}_3$ about a plaquette, that has been the focus of earlier proposals for experimental detection of spin liquid states~\cite{WWZ,ShastryShraiman,Chirality1,Chirality2}. We now review the standard hydrodynamics of the high-temperature phase.

\subsection{High-temperature hydrodynamics}

In the high-temperature, paramagnetic phase, the hydrodynamic modes correspond to the na{\"i}ve conservation laws of the model, namely energy and the three components of spin,
\begin{align}
\partial_t n_E + \nabla \cdot  \mathbf{j}_E &= 0,\\
\partial_t n_{S}^\alpha + \nabla \cdot \mathbf{j}^\alpha_{S}  &= 0, \quad \alpha=1,2,3.
\end{align}
In a given equilibrium state, the expectation values of the particle currents vanish, so that the linear response behaviour of the currents near 
equilibrium is given by the constitutive relations
\begin{align}
\label{eq:para1}
\delta \mathbf{j}_E &= -D_{EE}\nabla \delta n_E, \\
\label{eq:para2}
\delta \mathbf{j}_S^\alpha &= -D_{SS} \nabla \delta n^\alpha_S, \quad \alpha=1,2,3.
\end{align}
to leading order in a derivative expansion; these are decoupled by symmetry. Thus the paramagnetic phase exhibits four diffusive hydrodynamic modes in total, with dispersion
\begin{align}
i \omega_E &= D_{EE} k^2, \\
i\omega_{S^{\alpha}} &= D_{SS} k^2, \quad \alpha=1,2,3.
\end{align}

One might ask how far interaction effects, that appear in hydrodynamics as nonlinear corrections to the constitutive relations, can modify this na{\"i}ve picture of normal diffusion. For example, the leading nonlinear corrections to Eqs. \eqref{eq:para1} and \eqref{eq:para2} (note that simple total derivatives, e.g. $\delta n_E \nabla \delta n_E$, do not generate hydrodynamic tails~\cite{Husetails}) have the form
\begin{align}
\nonumber
\delta \mathbf{j}_S^\alpha &= -D_{SS} \nabla \delta n^\alpha_S - D^{(2)}_{SES} \delta n_E \nabla \delta n^{\alpha}_S+\ldots , \quad \alpha=1,2,3.
\end{align}
Such nonlinear corrections lead to long-time tails $\langle \mathcal{J}(t)\mathcal{J}\rangle \sim 1/t^{d/2+1}$ in the current autocorrelation functions of diffusing charges~\cite{Husetails} and tails $\sim 1/t^{d/2}$ for ballistic ones~\cite{spohn2016fluctuating}. In two dimensions, nonlinear diffusive corrections are nonsingular ($\propto \omega$) as $\omega \to 0^+$, while nonlinear ballistic corrections give rise to weak (logarithmic) finite-size dependence of transport coefficients~\cite{spohn2016fluctuating}. We will therefore disregard all such effects in the remainder of this paper, as they do not modify the long-time dynamical exponent measured in a given, finite, two-dimensional system.

We now examine how this simple picture of four, decoupled, diffusive hydrodynamic modes breaks down in spin liquid states at low temperature.

\section{Hydrodynamics of the $U(1)$ spin liquid with a Fermi surface}

At low temperatures, the square-lattice Heisenberg model is conjectured to support a zero-flux state~\cite{SLRMP}. The effective action for low-lying excitations about this state consists of two fermion (``spinon'') fields coupled to an emergent non-compact $U(1)$ gauge field, which can be written as a Lagrangian density\cite{Savary_2016,SLRMP}
\begin{align}
\nonumber \mathcal{L} = \sum_{\sigma = 1,2} &\psi_{\sigma}^\dagger(i \partial_t + e^* a_0) \psi_\sigma - \frac{1}{2m^*} \left|\left(\nabla - \frac{e^*}{c^*}i\mathbf{a}\right)\psi_\sigma\right|^2 \\
\label{eq:SLaction}
& +\frac{1}{8\pi}(e^2-b^2)+\ldots
\end{align}
to leading order in fields. For clarity, we have restored dimensions for the speed of light $c^*$ of the internal $U(1)$ field, as well as the effective mass $m^*$ and internal $U(1)$ charge $e^*$ of the spinon, but set $\hbar=1$. We shall also use vector notation $\mathbf{e}, \, \mathbf{b}$ for the Maxwell fields, where it is understood that $e_z = b_x=b_y=0$. Our immediate goal will be to derive the linear-response hydrodynamics of the effective action given by Eq. \eqref{eq:SLaction}.

\subsection{Regime of validity}
Let us first address the regime of validity of a hydrodynamic description based on the continuum field theory Eq. \eqref{eq:SLaction}, which turns out to be quite subtle. Superficially, the action Eq. \eqref{eq:MicHamiltonian} exhibits two symmetries that are not possessed by the underlying lattice model Eq. \eqref{eq:MicHamiltonian}, namely continuous translation symmetry and $U(1)$ gauge symmetry. At the level of hydrodynamics, these give rise to conservation of momentum $\int d^2 x \, \mathbf{P}$ and conservation of the total $U(1)$ flux, $\Phi = \int d^2 x \, b_z$. Both symmetries are approximate; even in the absence of disorder, the underlying lattice relaxes momentum on a timescale $\tau_{\mathbf{P}}$ associated with Umklapp scattering, while microscopic compactness of the emergent gauge field $\mathbf{a}$ leads to a violation of $U(1)_{\mathrm{flux}}$ symmetry~\cite{HermeleStab}, giving rise to fluctuations in the topological $U(1)_{\mathrm{flux}}$ quantum number $\Phi$ on some timescale $\tau_{\mathbf{a}}$ to be determined. For comparison with these timescales, it will be useful to introduce a scattering time $\tau_{\psi}$ for the spinon degrees of freedom.

In what follows, we shall focus on the regime $\tau_{\mathbf{a}}^{-1} \ll \omega \ll \tau_{\psi}^{-1}$. In this regime of frequencies, $U(1)_{\mathrm{flux}}$ symmetry holds approximately, so that the gauge field $\mathbf{a}$ may be modelled as non-compact, while the spinon degrees of freedom have sufficient time to reach a state of local equilibrium. The question now arises of whether such a regime exists at all, i.e. whether the condition $\tau_{\mathbf{a}}^{-1} \ll \tau_{\psi}^{-1}$ holds in some regime of temperature and disorder. The temperature dependence of these timescales is not well-studied. For concreteness, we focus on the clean case, for which it has been argued that $\tau_{\psi} \sim \tau_{\mathbf{P}} \sim T^{-4/3}$ due to scattering off the internal gauge degrees of freedom~\cite{RelRate1,RelRate2}. In particular, momentum is relaxed in the hydrodynamic regime.

We now put forward a Fermi Golden Rule argument for the temperature dependence of the timescale $\tau_{\mathbf{a}}$ for fluctuations in the $U(1)_{\mathrm{flux}}$ quantum number. Denote by $H_0$ the quantum Hamiltonian associated with the leading terms in the Lagrangian Eq. \eqref{eq:SLaction}. As written, the Lagrangian description pertains to a non-compact gauge field. Compactness of the gauge field can be restored by adding ``vortex creation'' operators
\begin{equation}
m^{\dagger}_{\mathbf{x}_0} = \exp{\left(i\int d^2 x\, \mathbf{a}^{\mathrm{cl}}_{\mathbf{x}_0} \cdot \mathbf{e}\right)}
\end{equation}
to the non-compact effective Hamiltonian $H_0$. Here, $\mathbf{a}^{\mathrm{cl}}_{\mathbf{x}_0}$ denotes a singular classical gauge texture with a $2\pi$-flux vortex centered at $\mathbf{x}_0$, so that by virtue of the canonical commutation relations $[a_i(\mathbf{x}),e_j(\mathbf{x}')] = i\delta_{ij}\delta(\mathbf{x}-\mathbf{x}')$, the operator $m^{\dagger}_{\mathbf{x}_0}$ realizes a vortex in the quantum gauge field $\mathbf{a}$. Notice that the operator $m^{\dagger}_{\mathbf{x}_0}$ is invariant under gauge transformations of $\mathbf{a}$, because the only operator appearing in its definition is the gauge-invariant quantum field $\mathbf{e}$. The definition of the classical gauge texture $\mathbf{a}^{\mathrm{cl}}_{\mathbf{x}_0}$ is not unambiguous, but the details beyond its $2\pi$ flux are not important~\cite{Pyrochlore}. Schematically, this gives rise to the expression
\begin{equation}
H = H_0 + g\int d^2 x_0 \cos{\left(\int d^2 x\, \mathbf{a}^{\mathrm{cl}}_{\mathbf{x}_0} \cdot \mathbf{e}\right)} + \ldots
\end{equation}
for the true Hamiltonian of the system. Thus we have included compactness of the gauge field as a perturbation to the effective Hamiltonian; this is entirely analogous to how compactness of the target space is encoded in $1+1$ dimensional Luttinger liquid theory, for example, when modelling Kosterlitz-Thouless-type transitions~\cite{giamarchi2004quantum}.

In the Euclidean time description, the operators $m^{\dagger}_{\mathbf{x}_0}$ create three-dimensional monopoles, as used by Polyakov to argue for confinement in compact $2+1$ dimensional Maxwell theory~\cite{POLYAKOV1977429}. In real time, such monopoles are correctly viewed as instantons, i.e. tunneling events that change the $U(1)_{\mathrm{flux}}$ quantum number of the state under consideration~\cite{HermeleStab}. The stability of the $U(1)$ spin liquid with a Fermi surface at non-zero temperature $T>0$ is contingent on irrelevance of the vortex creation operators $m^{\dagger}_{\mathbf{x}_0}$. Following various authors~\cite{SSLee,KSKim}, we shall assume that the vortex creation operator exhibits scaling with a highly irrelevant scaling dimension $d_{\mathbf{a}} \gg 3$. Neither assumption is trivial, as the spin liquid with a Fermi surface is neither scaling invariant nor known rigorously to be stable to confinement in two dimensions~\cite{Herbut,HermeleStab}; we refer to Ref.~\onlinecite{SSLee} for an argument in favour of both hypotheses. (We note that both assumptions are easier to justify for Dirac spin liquids~\cite{HermeleStab,ASL}, as discussed in Sec. \ref{sec:ASL} below.)

Subject to these assumptions, the timescale for fluctuations in $\Phi$ can be deduced as follows. By Fermi's Golden Rule, the timescale for instantons to generate scattering between eigenstates of $H_0$ scales with the coupling strength of the vortex creation operator, as $\tau_{\mathbf{a}} \sim g^{-2}$. This represents the shortest timescale on which the $U(1)_{\mathrm{flux}}$ quantum number in a thermal state of the effective Hamiltonian $H_0$ can vary due to instanton corrections. It remains to deduce the temperature dependence of $g$. To this end, let $g_0$ denote the bare value of the coupling $g$ at some cut-off scale $\Lambda \gg T$. At low temperature, the coupling $g$ is renormalized from its bare value $g_0$ by a factor $g = \left(\frac{T}{\Lambda}\right)^{d_\mathbf{a}}g_0$. It follows that $g\sim T^{d_\mathbf{a}}$, which yields
\begin{equation}
\tau_{\mathbf{a}} \sim T^{-2d_\mathbf{a}}.
\end{equation}
Strong irrelevance of the vortex creation operator $d_\mathbf{a} \gg 3$ then implies a clear separation of scales $\tau_{\mathbf{a}}^{-1} \ll \tau_{\psi}^{-1}$, giving rise to a hydrodynamic regime
\begin{equation}
\label{eq:TimeRegime}
\tau_{\mathbf{a}}^{-1} \sim T^{2 d_\mathbf{a}} \ll \omega \ll \tau_\psi^{-1} \sim T^{4/3}.
\end{equation}

\begin{figure*}[t]
    \centering
    \includegraphics[width=0.8\linewidth]{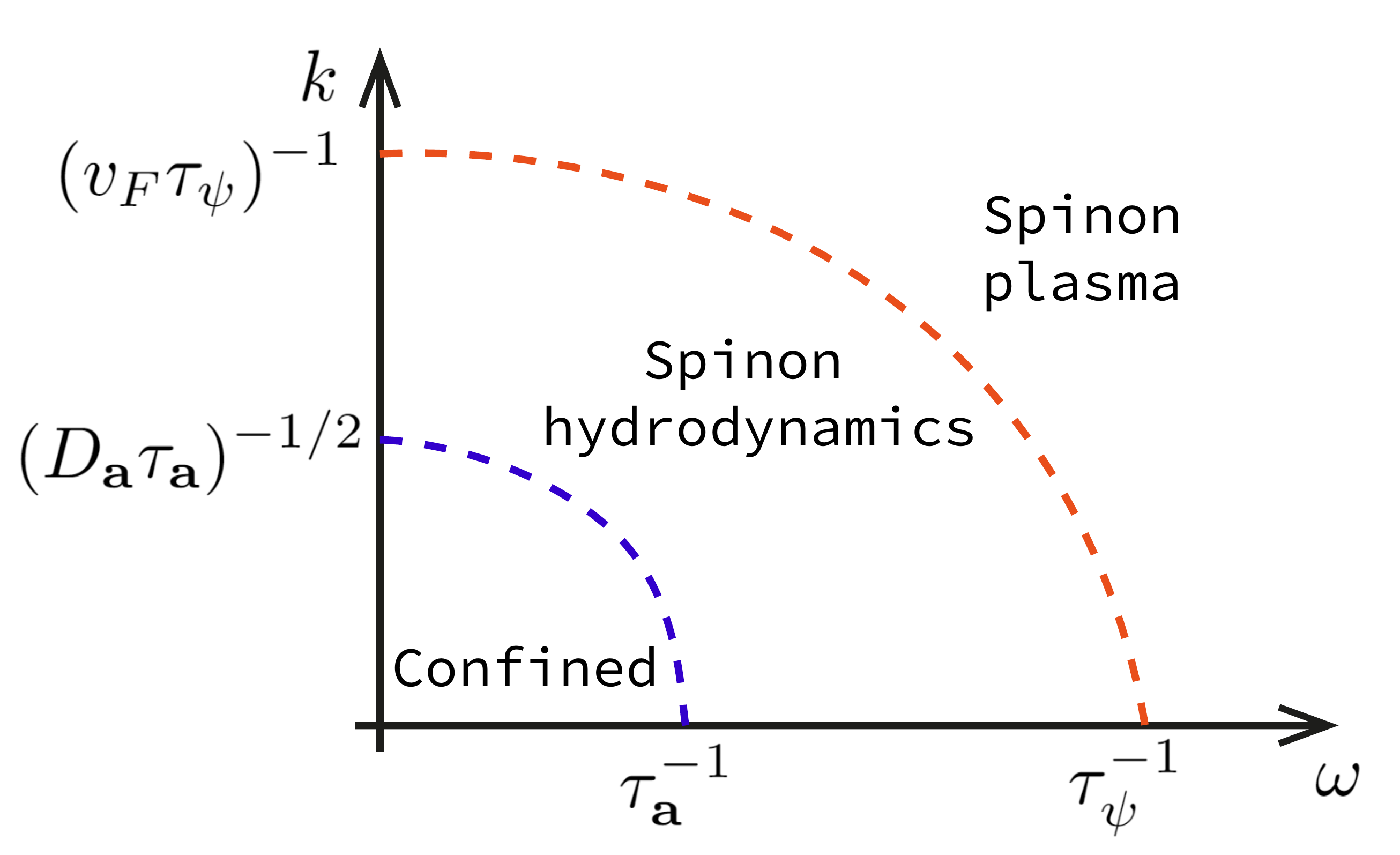}
    \caption{Schematic plot of the hydrodynamic regime in quantum spin liquids. The hydrodynamic frequency range is bounded on either side by the rate of fluctuations in the $U(1)_{\mathrm{flux}}$ quantum number $\tau_{\mathbf{a}}^{-1}$ and the rate of spinon scattering $\tau_{\psi}^{-1}$. At timescales smaller than $\tau_{\psi}$, the spinon degrees of freedom are no longer in equilibrium with each other, and the system is in a ``spinon plasma'' regime. At timescales larger than $\tau_{\mathbf{a}}$, compactness of the microscopic gauge field becomes apparent and leads to confinement. A magnetohydrodynamic regime of locally equilibrated spinon degrees of freedom coupled to an approximately non-compact gauge field arises between these two extremes. The origin of the associated wavenumber scales is discussed in the main text. The temperature dependence of these cutoffs is not well-studied in general and depends on the model at hand. For the $U(1)$ spin liquid with a Fermi surface, estimates are provided in Eqs. \eqref{eq:TimeRegime} and \eqref{eq:SpaceRegime}, while for Dirac spin liquids with a large number of fermion flavour $N_f \gg 1$, estimates are provided in Eqs. \eqref{eq:TimeRegimeASL} and \eqref{eq:SpaceRegimeASL}.}
    \label{Fig1}
\end{figure*}

Associated with these frequency cut-offs is a range of allowed wavenumbers. The microscopic motion of spinons is expected to be ballistic at a speed fixed by the Fermi velocity, leading to a microscopic length-scale $l_{\psi} = v_F \tau_{\psi}$ for spinon scattering. Meanwhile, on frequency scales $\tau_{\mathbf{a}}^{-1} \ll \omega \ll \tau_{\psi}^{-1}$, the internal gauge field is both locally equilibrated and approximately non-compact. This means that the magnetohydrodynamic description discussed below holds sway, indicating diffusive hydrodynamic transport of the transverse photon mode with a dynamical exponent $z=2$, according to Eq. \eqref{eq:transphot}. This allows us to self-consistently deduce a length-scale $l_{\mathbf{a}} = (D_{\mathbf{a}} \tau_{\mathbf{a}})^{1/2}$ below which the emergent magnetic field $\mathbb{b}$ exhibits hydrodynamic behaviour, where the diffusion constant $D_{\mathbf{a}}$ is defined below in Eq. \eqref{eq:photdiff}. Since we have assumed that microscopic scattering off the gauge field is the dominant contribution to spinon relaxation, we additionally have $D_{\mathbf{a}} \sim \sigma^{-1} \sim \tau_{\psi}^{-1} \sim T^{4/3}$, implying a hydrodynamic regime of wavenumbers
\begin{equation}
\label{eq:SpaceRegime}
(D_\mathbf{a} \tau_{\mathbf{a}})^{-1/2} \sim T^{d_\mathbf{a}-2/3} \ll k \ll (v_F\tau_\psi)^{-1} \sim T^{4/3}
\end{equation}
associated with the hydrodynamic regime of frequencies, Eq. \eqref{eq:TimeRegime}. Finally, we note that above the frequency and wavenumber scales $\tau_{\psi}^{-1}$ and $l_{\psi}^{-1}$, the spinon degrees of freedom do not have time to equilibrate with each other or with the internal gauge field. This defines a ``spinon plasma'' regime, whose behaviour is beyond the scope of this work (though see Ref. \cite{khoo2021universal} for a recent study that models this regime as a Fermi liquid). The range of possible dynamical regimes is summarized in Fig. \ref{Fig1}.

\subsection{Linear-response hydrodynamics}
In the regime of frequencies and wavenumbers given by Eqs. \eqref{eq:TimeRegime} and \eqref{eq:SpaceRegime}, spinon degrees of freedom are in local thermal equilibrium with each other and with microscopic fluctuations of the gauge field, while the gauge field itself can be modelled as non-compact. For non-zero temperatures $T>0$, this means that the effective Lagrangian Eq. \eqref{eq:SLaction} admits a classical magnetohydrodynamic regime largely analogous to the usual hydrodynamics of charged particles coupled to macroscopic Maxwell fields, except that momentum is no longer conserved.

We now derive the linear-response magnetohydrodynamics of Eq. \eqref{eq:SLaction}. First note that conservation of energy and spin yields four hydrodynamic modes,
\begin{align}
\label{eq:firstcons}
\partial_t n_E + \nabla \cdot  \mathbf{j}_E &= 0,\\
\partial_t n_{S}^\alpha + \nabla \cdot \mathbf{j}^\alpha_{S}  &= 0, \quad \alpha=1,2,3,
\end{align}
just as for the paramagnet. These must, however, be augmented by the $(2+1)$D Maxwell equations,
\begin{align}
\nabla \cdot \mathbf{b} &= 0,\\
\nabla \times \mathbf{e} &= -\frac{1}{c^*} \partial_t \mathbf{b}, \\
\nabla \cdot \mathbf{e} &= 4\pi n, \\
\nabla \times \mathbf{b} &= \frac{1}{c^*} \partial_t \mathbf{e} + \frac{4\pi}{c^*} \mathbf{j}.
\end{align}
Together, these imply conservation of the internal $U(1)$ charge (spinon density)
\begin{align}
\label{eq:lastcons}
\partial_t n + \nabla \cdot \mathbf{j} &= 0.
\end{align}
In the interests of treating gauge and spinon degrees of freedom on equal footing, we have combined ``field'' and ``matter'' contributions in the energy density $n_E$. The advantage of this approach is that effects such as Joule heating, which consists of a transfer of energy from fields to matter but does not violate local energy conservation, are taken into account automatically. Note also that because momentum is relaxed on the hydrodynamic timescale, $\tau_{\mathbf{P}}^{-1} \sim \tau_{\psi}^{-1}$, there are no sound modes (in contrast to the Dirac spin liquid, as discussed in Sec. \ref{sec:ASL} below).

Let us restrict to the linear response regime, and study small perturbations $\mathbf{e} = \mathbf{e}_0+\delta \mathbf{e}$ and $n = n_0 + \delta n$ about a stationary background charge distribution $\nabla \cdot \mathbf{e}_0 = 4\pi n_0$. We assume general, isotropic, linear-response constitutive relations of the form
\begin{align}
\label{eq:const1}
\delta \mathbf{j}_S^\alpha &= -D_{SS} \nabla \delta n^\alpha_S, \\
\label{eq:const2}
\delta \mathbf{j} &= \sigma \delta \mathbf{e} - D_{nn}\nabla \delta n - D_{nE}\nabla \delta n_E, \\
\delta \mathbf{j}_E &= \sigma' \delta \mathbf{e} - D_{En}\nabla \delta n - D_{EE}\nabla \delta n_E
\end{align}
Note that the local magnetization is decoupled from the other hydrodynamic modes at linear order, since no other gradients transform appropriately under spin rotation. Similarly, vector currents cannot couple to gradients of the pseudoscalar $b_z$ at linear order.

We find five hydrodynamic modes in total (see Appendix for details). There are three diffusive magnetization modes,
\begin{equation}
\label{eq:u1spin}
i\omega_{S^{\alpha}} = D_{SS} k^2, \quad \alpha=1,2,3,
\end{equation}
as in the paramagnetic phase. There is again a diffusive energy mode
\begin{equation}
\label{eq:diffE}
i\omega_E = \left(D_{EE}-\frac{\sigma'}{\sigma} D_{nE}\right)k^2 + \mathcal{O}(k^4)
\end{equation}
corresponding to fluctuations of $\delta n_E$. However, in the spin liquid phase, this mode is accompanied by fluctuations of the longitudinal slow modes $\{\delta n,\delta \mathbf{e}_{\parallel},\delta \mathbf{j}_{\parallel},\delta \mathbf{j}_{E,\parallel}\}$.

Finally there is an additional hydrodynamic mode compared to the paramagnet, which arises due to the emergent $U(1)$ gauge symmetry of the spin liquid phase. This is the transverse photon mode,
\begin{equation}
\label{eq:transphot}
i\omega_{\mathbf{a}} = D_{\mathbf{a}}k^2 + \mathcal{O}(k^4),
\end{equation}
which in the presence of a non-zero conductivity $\sigma>0$ is \emph{diffusive} at long wavelengths, with effective diffusion constant
\begin{equation}
\label{eq:photdiff}
D_{\mathbf{a}} = \frac{{c^*}^2}{4\pi \sigma}.
\end{equation}
The transverse photon is an excitation of the slow modes $\{\delta \mathbf{e}_{\perp}, \delta \mathbf{b}, \delta \mathbf{j}_{\perp}, \delta \mathbf{j}_{E,\perp}\}$. Upon fixing $\delta \mathbf{e}_{\perp}$, the induced magnetic field fluctuations $\delta \mathbf{b} = (c^*/\omega) \mathbf{k} \times \delta \mathbf{e}_{\perp}$ by Faraday's law, while the transverse current fluctuations are proportional to $\delta \mathbf{e}_\perp$. The main point is that the transverse photon can be excited by driving fluctuations of the internal magnetic field $\mathbf{b}$. The latter is related to the microscopic spin chirality~\cite{WWZ} $\mathcal{P} = \mathbf{S}_1 \cdot \mathbf{S}_2 \times \mathbf{S}_3$ about a plaquette. This might provide one physical route towards coupling to the hydrodynamic transverse photon mode~\cite{ShastryShraiman,Chirality1,Chirality2}. 

This mode can be viewed as the spin-liquid analogue of magnetic diffusion that arises in ordinary conductors~\cite{jackson1999classical}. In more detail, the hydrodynamic interpretation of the ``skin effect'' for electromagnetic fields in an electrical conductor is that it describes a crossover from ballistic to diffusive behaviour~\cite{BAGGIOLI20201}. In the present case, the crossover scale $k^*$ such that diffusive behaviour sets in for $k \ll k^*$ is given by (see Appendix) $k^*  = 2\pi \sigma / c^* \sim \tau_{\psi} \sim T^{-4/3}$, which lies well beyond the UV cutoff in Eq. \eqref{eq:SpaceRegime} at low temperatures. Thus the long-wavelength hydrodynamics of the transverse internal photon is diffusive, rather than ballistic as one might na{\"i}vely expect.

We conclude this section with some speculative remarks on the hydrodynamics of the spinon Fermi surface with conserved momentum. Standard treatments~\cite{RelRate1,RelRate2}, based on physically realistic lattice geometries that are conjectured to yield a spinon Fermi surface state, find that $\tau_{\mathbf{P}}$ and $\tau_{\psi}$ are the same order of magnitude, so that momentum is always relaxed in the hydrodynamic regime. However, the effective action Eq. \eqref{eq:SLaction} is meaningful for Fermi surfaces that do not relax momentum efficiently. For example, a suppression of Umklapp scattering would imply a momentum relaxation rate exponentially large in inverse temperature~\cite{RoschAndrei,Shimshoni}, so that $\tau_{\mathbf{P}}^{-1} \ll \tau_{\mathbf{a}}^{-1}$. In this case, one would expect ballistic sound modes of charge, momentum and energy on generic grounds~\cite{hartnoll2018holographic}, but the usual analysis is complicated by the possibility of momentum exchange between field and matter degrees of freedom via Poynting flux. In particular, it is not clear whether a ballistically propagating charge mode is physical; a definitive resolution of this question seems to lie beyond linear-response hydrodynamics.

\section{Hydrodynamics of the algebraic spin liquid}
\label{sec:ASL}
As a more sophisticated illustration of the hydrodynamic approach, we now consider the hydrodynamics of the $SU(4)$-symmetric ``algebraic spin liquid" that arises in the staggered flux phase of the square-lattice Heisenberg model. The low-energy effective action for this state can be expressed as the Lagrangian density~\cite{ASL}
\begin{equation}
\label{ASLaction}
\mathcal{L} = \overline{\Psi}\left[-i\gamma^\mu\left(\partial_{\mu}+\frac{ie^*}{c^*}a_\mu\right)\right] \Psi + \frac{1}{8\pi}(e^2-b^2)+\ldots
\end{equation}
where $\Psi$ denotes an eight-component spinor,
\begin{equation}
\Psi = \begin{pmatrix} \psi_{11} \\ \psi_{12} \\ \psi_{21} \\ \psi_{22} \end{pmatrix},
\end{equation}
in which each $\psi_{ia}$ is a two-component spinor, $i$ denotes a valley index and $a$ is the spin index. Following Ref.~\onlinecite{ASL}, we denote Pauli matrices acting on the Dirac, valley and spin degrees of freedom by $\{\tau^i,\mu^i,\sigma^i\}$ respectively, and define $\overline{\Psi} = i\Psi^\dagger\tau^3$. The $U(1)$ internal and gauge symmetries act similarly to the uniform flux phase considered above. However, the $SU(2)$ spin symmetry now sits within a larger, internal $SU(4)$ symmetry flavour group, that is generated by the fifteen operators $\{\sigma^i, \mu^i,\sigma^i\mu^j\}$. 

\subsection{Regime of validity}
As above, the regime of validity of the hydrodynamic description we propose is bounded on either side by the scattering rates for instanton and spinon degrees of freedom, with the separation of scales $\tau_{\mathbf{a}}^{-1} \ll \tau_{\psi}^{-1}$ a necessary condition for its existence. By emergent Lorentz invariance and the presence of strong spinon-gauge interactions, the spinon scattering rate is expected to exhibit Planckian scaling~\cite{QCP} $\tau_{\psi}^{-1} \sim T$. We again estimate the timescale for fluctuations of the $U(1)_{\mathrm{flux}}$ quantum number as $\tau_{\mathbf{a}} \sim T^{-2d_{\mathbf{a}}}$, but for the algebraic spin liquid with a small number of fermion flavours $N_f=4$, as in Eq, \eqref{ASLaction}, it is not known whether instantons are irrelevant~\cite{ASL}, i.e. whether $d_{\mathbf{a}} > 3$. For a large number of fermion flavours $N_f \gg 1$, it is expected~\cite{HermeleStab} that $d_{\mathbf{a}} = \mathcal{O}(N_f)$, so that a hydrodynamic regime
\begin{equation}
\label{eq:TimeRegimeASL}
\tau_{\mathbf{a}}^{-1} \sim T^{\mathcal{O}(N_f)} \ll \omega \ll \tau_{\psi}^{-1} \sim T
\end{equation}
exists. The associated length scales may be determined as above. On hydrodynamic timescales $\omega \ll \tau_{\psi}^{-1} \sim T$, the gauge degrees of freedom are again expected to propagate diffusively with dynamical exponent $z=2$, leading to a length scale $l_{\mathbf{a}} \sim (D_\mathbf{a} \tau_{\mathbf{a}})^{1/2} \sim T^{1/2-d_\mathbf{a}} \sim T^{-\mathcal{O}(N_f)}$ in this case. Meanwhile, the microscopic motion of spinons is again expected to be ballistic with speed $v_F = c^*$, leading to a hydrodynamic regime of wavenumbers
\begin{equation}
\label{eq:SpaceRegimeASL}
(D_\mathbf{a} \tau_{\mathbf{a}})^{-1/2} \sim T^{\mathcal{O}(N_f)} \ll k \ll (c^*\tau_\psi)^{-1} \sim T.
\end{equation}

Thus we expect a robust magnetohydrodynamic regime in the limit of a large number of fermion flavours, as depicted in Fig. \ref{Fig1}. The existence of this regime is less clear for $N_f=4$; nevertheless, the basic $N_f=4$ case serves as an instructive template for the hydrodynamics of spin liquid states with higher flavour symmetry and emergent Lorentz invariance.

Finally, we note that the continuum Lagrangian Eq. \eqref{ASLaction} is translationally invariant although the underlying lattice model is not. Since there is no Fermi surface, however, the lowest-order processes in the lattice model that relax momentum are Umklapp scattering events between Fermi points, which cost an energy $E_U \sim v_F |\mathbf{G}|$, where $|\mathbf{G}| = 2\pi/a$ is the magnitude of a reciprocal lattice vector on the square lattice with lattice spacing $a$. Thus for a clean system, the momentum scattering time $\tau_{\mathbf{P}} \sim e^{E_U/k_BT}$ is exponentially large at low temperatures~\cite{RoschAndrei,Shimshoni}, and momentum must be included as a conserved mode on the timescales of Eq. \eqref{eq:TimeRegimeASL}.

\subsection{Linear-response hydrodynamics}

The hydrodynamics of the algebraic spin liquid can again be modelled by the equations Eq. \eqref{eq:firstcons}-\eqref{eq:lastcons} introduced for the $U(1)$ spin liquid with a Fermi surface, i.e. the Maxwell equations and conservation of energy and spin. However, there are several important differences compared to the Fermi surface case. First, since $\tau_{\mathbf{P}}^{-1} \ll \tau_{\mathbf{a}}^{-1}$ due to microscopic suppression of Umklapp scattering, momentum must be treated as a conserved mode in the putative hydrodynamic regime. This implies an additional hydrodynamic equation
\begin{equation}
\label{eq:momcons}
\partial_t P_i + \partial_j \Pi_{ij} = 0, \quad i,j=x,y,
\end{equation}
where $\mathbf{P}=(P_x,P_y)$ is the local momentum density and $\Pi$ denotes the momentum flux tensor. (As for the energy density $n_E$ in the previous section, we have implicitly combined ``field'' and ``matter'' contributions in the definitions of $P_i$ and $\Pi_{ij}$.) Second, the effective Lagrangian Eq. \eqref{ASLaction} is Lorentz invariant. This means that in the limit that lattice effects are neglected, the equations of motion associated with Eq. \eqref{ASLaction} are Lorentz invariant. Finally, there are twelve additional flavour modes corresponding to the internal $SU(4)$ symmetry. However, these transform non-trivially under elements of the lattice space group and are therefore decoupled from the slow modes of Eqs. \eqref{eq:firstcons} - \eqref{eq:lastcons} and \eqref{eq:momcons} at the level of linear response, and may be considered separately.

The modifications due to emergent Lorentz symmetry can be derived as follows. First notice that linear dissipative corrections include lattice effects and are therefore exempt from the requirement of Lorentz covariance. In fact, the only substantive consequence of emergent Lorentz symmetry is that the energy and momentum operators must transform as a tensor in the limit that lattice effects are neglected. In this limit, the operator equations of motion for energy and momentum associated with Eq. \eqref{ASLaction} take the form
\begin{align}
\partial_t \hat{n}_E + {c^{*}}^2  \partial_j \hat{P}_j &= 0, \\
\partial_t \hat{P}_i + \partial_j \hat{\Pi}_{ij} &= 0.
\end{align}
Hydrodynamic equations at the Euler (ballistic) scale are obtained by averaging the operator equations of motion over locally thermal states, and here yield the constitutive relation
\begin{equation}
\mathbf{j}_E = {c^*}^2 \mathbf{P}
\end{equation}
for the energy current. Restoring Navier-Stokes (diffusive) corrections and passing to the linear response regime, it follows that the linear-response energy current picks up a reactive coupling to momentum
\begin{equation}
\label{eq:modconst}
\delta \mathbf{j}_E = {c^*}^2 \delta \mathbf{P} + \sigma' \delta \mathbf{e} - D_{En}\nabla \delta n - D_{EE}\nabla \delta n_E,
\end{equation}
while the linear-response momentum current is given by
\begin{align}
\label{eq:const4}
\nonumber \delta \Pi_{ij} &= \left(\frac{\partial{p}}{\partial n}\delta n +\frac{\partial{p}}{\partial n_E}\delta n_E\right)\delta_{ij} \\
& - \nu (\partial_i \delta P_j + \partial_j \delta P_i - \delta_{ij} \nabla \cdot \delta \mathbf{P}) - \zeta \delta_{ij} \nabla \cdot \delta \mathbf{P}.
\end{align}
where $p(n,n_E)$ is the equilibrium pressure and $\nu$ and $\zeta$ define kinematic shear and bulk viscosities respectively. The generic form of the other constitutive relations, Eqs. \eqref{eq:const1} and \eqref{eq:const2}, is unchanged.

We deduce that of the five hydrodynamic modes present for the $U(1)$ spinon Fermi surface state, four are unchanged for the algebraic spin liquid, namely the three components of spin and the transverse photon mode
\begin{align}
i\omega_{S^{\alpha}} &= D_{SS} k^2, \quad \alpha=1,2,3,\\
i\omega_{\mathbf{a}} &= D_{\mathbf{a}}k^2 + \mathcal{O}(k^4).
\end{align}
Notice that at low temperatures, the wavenumber scale $k^* = 2\pi \sigma/c^* \sim T^{-1}$ below which the internal transverse photon crosses over from ballistic to diffusive behaviour is once again well beyond the UV cutoff for the hydrodynamic regime of wavenumbers Eq. \eqref{eq:SpaceRegimeASL}.

By virtue of the reactive coupling in Eq. \eqref{eq:modconst}, the energy mode combines with the longitudinal momentum mode to yield a ballistic sound mode for the algebraic spin liquid, at least on the timescales $t \ll \tau_{\mathbf{P}}$ for which momentum appears conserved. This phenomenon has been noted in other condensed matter systems with emergent Lorentz symmetry, such as graphene~\cite{Levitov,LucasSound}. The result (see Appendix) is that the energy and longitudinal momentum modes combine to form a ``sound mode'' with diffusive broadening,
\begin{equation}
\omega_{s} = \pm c_s k + \mathcal{O}(k^2),
\end{equation}
where the effective speed of sound $c_s = \sqrt{\frac{\partial p}{\partial n_E}} c^*$. If we further assume an ultrarelativistic, two-dimensional equation of state $p=n_E/2$, on the grounds that the effective action Eq. \eqref{ASLaction} describes purely massless excitations, we recover the universal relation for sound modes in two-dimensional ultrarelativistic plasma~\cite{Levitov,Kovtun_2012,Lucas_2018},
\begin{equation}
c_s = \frac{c^{*}}{\sqrt{2}}.
\end{equation}

These modes correspond to joint oscillations of the full set of longitudinal hydrodynamic modes $\{\delta n_E, \delta n, \delta \mathbf{e}_{\parallel}, \delta \mathbf{j}_{\parallel}, \delta \mathbf{j}_{E,\parallel}, \delta \mathbf{P}_{\parallel}\}$. We emphasize that ballistic behaviour of these modes is, strictly speaking, a property of the field theory Eq. \eqref{ASLaction} and not the underlying lattice model; in realistic settings, such proximate Lorentz invariance yields a vanishing Drude weight and an anomalously large thermal conductivity~\cite{RoschAndrei,Shimshoni}. It is nevertheless consistent to treat these modes as ballistic because the crossover timescale to diffusive behaviour, which is the momentum relaxation time $\tau_{\mathbf{P}}$, exceeds the IR cutoff of our hydrodynamic theory in Eq. \eqref{eq:TimeRegimeASL}, at least for clean systems in which $\tau_{\mathbf{P}} \sim e^{E_U/k_BT} \gg \tau_{\psi} \sim T^{-1}$.

Finally there is a diffusive, transverse momentum mode due to the shear viscosity, with dispersion relation
\begin{equation}
i\omega^{\perp}_{\mathbf{P}} = \nu k^2,
\end{equation}
that is decoupled from the other hydrodynamic modes.

We now turn to the twelve flavour modes that arise from $SU(4)$ symmetry, excluding the three spin modes considered above. As noted above, these twelve modes transform nontrivially under elements of the lattice space group, which prevents them from coupling linearly to any of the hydrodynamic modes considered so far. We first define flavour charge densities
\begin{align}
n_F^{i} &= \langle \Psi^\dagger \mu^{i} \Psi \rangle, \\
n_F^{i,\alpha} &= \langle \Psi^\dagger \sigma^\alpha \mu^{i} \Psi \rangle,
\end{align}
through local equilibrium averages of local observables, and associated current densities $\mathbf{j}_F^i,\, \mathbf{j}_F^{i,\alpha}$, which satisfy
\begin{align}
\partial_t n_F^i + \nabla \cdot \mathbf{j}_F^i &= 0, \quad i =1,2,3, \\
\partial_t n_F^{i,\alpha} + \nabla \cdot \mathbf{j}_F^{i,\alpha} &= 0, \quad i,\alpha=1,2,3.
\end{align}
These conserved densities correspond to competing order parameters of the algebraic spin liquid phase, each of which arises from a generator of the $SU(4)$ flavour symmetry~\cite{ASL}. For example, $n_F^{1/2}$ acts as an order parameter for valence-bond-solid (VBS) order, while $n_F^3$ is a local proxy for the density of N{\'e}el Skyrmions.

Now consider the linear-response hydrodynamic behaviour about a fully disordered equilibrium state, in which $n_{F,0}^i = n_{F,0}^{i,\alpha}=0$ and all such competing orders are on equal footing. The linear response of the system is stratified into symmetry sectors by the action of the lattice space group, yielding general constitutive relations
\begin{align}
\label{eq:firstj}
\delta \mathbf{j}_F^{1} &= - D_{11} \nabla \delta n^{1}_F - D_{12} \nabla \delta n^2_F, \\
\delta \mathbf{j}_F^{2} &= - D_{21} \nabla \delta n^{1}_F - D_{22} \nabla \delta n^2_F, \\
\delta \mathbf{j}_F^{3} &= - D_{33} \nabla \delta n^{3}_F, \\
\delta \mathbf{j}_F^{1,\alpha} &= - D'_{11} \nabla \delta n^{1,\alpha}_F - D'_{12} \nabla \delta n^{2,\alpha}_F, \\
\delta \mathbf{j}_F^{2,\alpha} &= - D'_{21} \nabla \delta n^{1,\alpha}_F - D'_{22} \nabla \delta n^{2,\alpha}_F, \\
\label{eq:lastj}
\delta \mathbf{j}_F^{3,\alpha} &= - D'_{33} \nabla \delta n^{3,\alpha}_F.
\end{align}
We have already exploited spin rotation symmetry of the background equilibrium state to reduce the number of independent linear-response coefficients to ten. In principle, the system Eqs. \eqref{eq:firstj}-\eqref{eq:lastj} is sensitive to VBS-type order of the background equilibrium state. If such order is absent, as we have assumed, then the diffusion tensors $\hat{D}=\begin{pmatrix} D_{11} & D_{12} \\ D_{21} & D_{22} \end{pmatrix}$, $\hat{D}' = \begin{pmatrix} D'_{11} & D'_{12} \\ D'_{21} & D'_{22} \end{pmatrix}$
in the VBS sector cannot pick out a preferred direction in the space of modes, yielding additional constraints $D_{21}=D_{12}=0$, $D_{11}=D_{22}$, $D'_{12}=D'_{21}=0$ and $D'_{11}=D'_{22}$, leaving only four independent transport coefficients, $\{D_{11}, D_{33}, D'_{11}, D'_{33}\}$. The resulting hydrodynamic normal modes are diffusive, and tabulated in Table \ref{tab:table}. 
\begin{table}[h]
\begin{tabular}{c|c|c}
    Mode & Dispersion & Multiplicity \\
    \hline
     $\delta n_F^{1},\,\delta n_F^{2}$ & $\omega = -iD_{11}k^2$ & 2 \\
     $\delta n_F^{3}$ & $\omega = -iD_{33}k^2$ & 1 \\
     $\delta n_F^{1,\alpha},\delta n_F^{2,\alpha}$ & $\omega = -iD'_{11}k^2$ & 6 \\
     $\delta n_F^{3,\alpha}$ & $\omega = -iD'_{33}k^2$ & 3 \\
\end{tabular}
\caption{Hydrodynamic normal modes of the system Eqs. \eqref{eq:firstj}-\eqref{eq:lastj}.}
\label{tab:table}
\end{table}

These multiplicities are in agreement with the lattice symmetry classification of the $SU(4)$ flavour generators that was given in previous work~\cite{ASL}. We also note that a one-to-one correspondence between the Noether charges arising from internal non-Abelian symmetries and diffusive hydrodynamic modes was derived recently~\cite{glorioso} as a prediction of mode-coupling theory, and is consistent with our treatment of internal symmetries above.

To summarize, the $SU(4)$-symmetric algebraic spin liquid state has nineteen hydrodynamic modes, of which fifteen arise from the internal $SU(4)$ flavour symmetry (this includes the three components of spin), one is the transverse photon mode discussed above for the spinon Fermi surface state, one is the diffusive transverse momentum mode, and two are ballistic, longitudinal energy-momentum modes arising from emergent Lorentz symmetry.

\section{Discussion}

Arguing based on general principles of symmetry and linear response, we have derived the linear-response hydrodynamics of two representative quantum spin liquid states. We find that the intricate low energy physics of such states, as made manifest by the gauge and flavour symmetries of their effective field theories, leads to a variety of hydrodynamic modes beyond the quartet of diffusive spin and energy modes that characterize ordinary paramagnets. 

Possible experimental signatures of nonlinear spinon physics have been the focus of recent theoretical attention~\cite{BalentsStarykh,Laumann}. Above, we identified a simple and generic feature of spin liquids' \emph{linear} response, namely the presence of a diffusive transverse mode of the internal gauge photon. Physically speaking, our results on the hydrodynamic modes arising from gauge and flavour symmetry yield predictions for the finite-temperature dynamics of various autocorrelation functions of local two-, three- and four-spin operators~\cite{WWZ,ASL} in the ``hydrodynamic'' range of frequencies and wavenumbers depicted in Fig. \ref{Fig1}. While the existence of such a hydrodynamic regime for spin liquids is relatively easy to justify in three spatial dimensions, for which compactness of the emergent gauge field does not generically destabilize the spin liquid state~\cite{HermeleStab}, its existence for physically realistic examples of spin liquid states in two dimensions is unclear. Pending a more rigorous understanding of the stability of two-dimensional spin liquid states, the arguments summarized in Fig. \ref{Fig1} predict a robust hydrodynamic regime at small $T>0$ for the two-dimensional Dirac spin liquid with a large number of fermion flavours~\cite{HermeleStab,ASL}, and possibly even for the $U(1)$ spin liquid with a Fermi surface~\cite{KSKim,SSLee}.

We emphasize that conclusive experimental evidence for $U(1)$ spin liquid phases in two dimensions remains beyond reach at present~\cite{Savary_2016,SLRMP}. So far, there only exist promising candidate materials that exhibit no magnetic ordering down to temperatures of the order of tens of mK, for example~\cite{Savary_2016,SLRMP} the triangular lattice materials $\mathrm{EtMe_3Sb[Pd(dmit)_2]_2}$ and $\mathrm{\kappa-(ET)_2Cu_2(CN)_3}$, and the kagome material $\mathrm{ZnCu_3(OH)_6Cl_2}$, also known as herbertsmithite. Thus the question arises of whether the hydrodynamic modes we identified above can plausibly be seen in such materials. A particularly well-studied candidate system for exploring this question is herbertsmithite, whose ground state is conjectured to be an algebraic spin liquid~\cite{HerbSmith1,HerbSmith2} (although it is difficult to rule out valence-bond solid order at the lowest temperatures\cite{Hastings}). For the conjectured spin liquid state in this system, it has been argued that the kagome lattice structure enables coupling of the scalar spin chirality to both Raman scattering~\cite{ShastryShraiman,KagomeRaman} and neutron scattering~\cite{Chirality2} measurements, and that both these measurements probe the autocorrelation function of the gauge magnetic field at zero temperature, which has singular features near ballistic dispersion $\omega \approx v_F |k|$. However, since hydrodynamics pertains to non-zero temperatures, these results are not directly applicable to the hydrodynamic regime. On this point, we note that inelastic neutron scattering has recently provided a convincing experimental demonstration of the dynamical exponents arising from hydrodynamics at non-zero temperature in quantum spin chains~\cite{KPZ}. This raises the possibility that for spin liquid states in kagome materials\cite{Chirality2}, neutron scattering measurements could similarly reveal the emergence of a diffusive gauge mode at small, non-zero temperatures, in addition to the three diffusive magnetization modes expected at all temperatures.

Our study of the hydrodynamics of quantum spin liquids connects to an ongoing theoretical effort to understand the implications of Lie group symmetry within hydrodynamics. Various anomalous transport phenomena in one-dimensional quantum systems can be traced to the presence of Lie group symmetry~\cite{Znidaric,IlievskiLattice,DupontMoore,gauge,ilievski2020superuniversality}, and it would be interesting to understand how far ideas useful in understanding spin liquid physics, for example, one-dimensional versions of parton constructions~\cite{Mudry1,Mudry2}, might shed light on these phenomena. In the context of effective field theory approaches to hydrodynamics~\cite{Crossley2017}, the spin liquid states studied in this paper motivate an extension of the hydrodynamic formalism recently developed for physical systems with global Lie group symmetries~\cite{glorioso} to systems with gauge symmetries. 

\textit{Acknowledgments.} We thank Igor Herbut and especially Andrew Lucas and Joseph Maciejko for several illuminating discussions. SLS acknowledges support from the United States Department  of  Energy  via  grant  No.   DE-SC0016244.

\bibliography{biblio.bib}

\onecolumngrid
\appendix

\section{Linear-response hydrodynamics of quantum spin liquids}
\subsection{Hydrodynamic modes of the $U(1)$ spin liquid with a Fermi surface}
\subsubsection{Hydrodynamic equations}
We first Fourier transform in space and time. It is helpful to write $\mathbf{v} = \mathbf{v}_{\parallel} + \mathbf{v}_{\perp}$ for vector components parallel and perpendicular to $\mathbf{k}$. In this notation, the Fourier transformed Maxwell equations read
\begin{align}
\delta \mathbf{b}_{\parallel} &= 0,\\
\label{eq:faraday}
\delta \mathbf{b}_{\perp} &= \frac{c^*}{\omega} \mathbf{k} \times \delta \mathbf{e}_{\perp},\\
\label{eq:gauss}
i \mathbf{k} \cdot \mathbf{e}_{\parallel} &= 4 \pi \delta n, \\
\mathbf{e}_{\parallel} &= \frac{4\pi}{i\omega}\mathbf{j}_{\parallel},\\
\label{eq:diel}
\left(k^2-\frac{\omega^2}{{c^*}^2}\right)\delta \mathbf{e}_{\perp} &= \frac{4\pi i \omega}{{c^*}^2} \delta \mathbf{j}_{\perp}.
\end{align}
Meanwhile, the constitutive relations for charge and energy current read
\begin{align}
\label{eq:jperp}
\delta \mathbf{j}_{\perp} &= \sigma_{\perp} \delta \mathbf{e}_{\perp} , \\
\label{eq:jparr}
\delta \mathbf{j}_{\parallel} &= \sigma_{\parallel} \delta \mathbf{e}_{\parallel}- i D_{nE} \delta n_E \mathbf{k}, \\
\label{eq:jeperp}
\delta \mathbf{j}_{E,\perp} &= \sigma'_{\perp} \delta \mathbf{e}_{\perp}, \\
\label{eq:jeparr}
\delta \mathbf{j}_{E,\parallel} &= \sigma'_{\parallel} \delta \mathbf{e}_{\parallel} - i D_{EE} \delta n_E \mathbf{k},
\end{align}
with transverse and longitudinal conductivities given by
\begin{align}
\sigma_{\perp} &= \sigma,\quad \sigma_{\parallel} = \sigma + \frac{1}{4\pi} D_{nn} k^2, \\
\sigma'_{\perp} &= \sigma',\quad \sigma'_{\parallel} = \sigma' + \frac{1}{4\pi} D_{En} k^2.
\end{align}

\subsubsection{Transverse hydrodynamic modes}
There is a single transverse hydrodynamic mode corresponding to the transverse photon of the internal gauge field. By the constitutive relations given in Eqs. \eqref{eq:jperp}-\eqref{eq:jeparr} and Eq. \eqref{eq:diel}, we deduce that transverse electric field excitations satisfy the equation
\begin{equation}
\label{eq:diel2}
\left[k^2 - \frac{\omega^2}{{c^*}^2}\left(1+\frac{4\pi i}{\omega} \sigma_{\perp}\right)\right] \delta \mathbf{e}_{\perp}=0,
\end{equation}
Eq. \eqref{eq:diel2} is the vanishing of the dielectric function familiar in ordinary conductors, and we deduce that a transverse photon propagates when the condition
\begin{equation}
1 + \frac{4\pi i \sigma}{\omega} - \frac{{c^*}^2k^2}{\omega^2}=0
\end{equation}
is satisfied, which has roots
\begin{equation}
\label{eq:transroots}
\omega = - 2 \pi i \sigma \pm \sqrt{{c^*}^2 k^2 - (2\pi \sigma)^2}.
\end{equation}

Only the positive branch of this equation is hydrodynamic (i.e. $\omega \to 0$ as $k\to 0$), and yields the dispersion relation
\begin{equation}
\omega_{\mathbf{a}} = -2 \pi i \sigma +  \sqrt{{c^*}^2 k^2 - (2\pi \sigma)^2}.
\end{equation}
This dispersion relation exhibits the well-known peculiarity of a ``k-gap", or skin effect\cite{BAGGIOLI20201}, due to its branch point at the critical wavenumber
\begin{equation}
k^* = \frac{2\pi \sigma}{c^*}.
\end{equation}
Thus short-wavelength excitations $k \gg k^*$ look like ballistic photon modes with $\omega \sim c^* k$, while the hydrodynamic regime, $k \ll k^*$, exhibits \emph{diffusive} behaviour with
\begin{equation}
\omega_{\mathbf{a}} \sim -i D_{\mathbf{a}}k^2 + \mathcal{O}(k^4), \quad k \to 0,
\end{equation}
where the photon diffusion constant
\begin{equation}
D_{\mathbf{a}} = \frac{{c^*}^2}{4\pi \sigma}.
\end{equation}

This diffusive transverse electric field mode induces magnetic fluctuations $\delta \mathbf{b} = (c^*/\omega)\mathbf{k} \times \delta \mathbf{e}_{\perp}$ by Faraday's law Eq. \eqref{eq:faraday}. By the constitutive relations Eqs. \eqref{eq:jperp} and \eqref{eq:jeperp} it is also accompanied by transverse charge and energy currents, $\delta \mathbf{j}_\perp = \sigma_\perp \delta \mathbf{e}_\perp$ and $\delta \mathbf{j}_{E,\perp} = \sigma'_\perp \delta \mathbf{e}_\perp$. Thus the transverse photon is a simultaneous excitation of the modes $\{\delta \mathbf{e}_{\perp}, \delta \mathbf{b}, \delta \mathbf{j}_{\perp}, \delta \mathbf{j}_{E,\perp}\}$.

\subsubsection{Longitudinal modes}
With the aid of Gauss's law Eq. \eqref{eq:gauss} and the constitutive relations Eqs. \eqref{eq:jperp}-\eqref{eq:jeparr}, the conservation laws for charge and energy density reduce to the coupled linear system
\begin{align}
\label{eq:longmodes}
\begin{pmatrix} i \omega - 4 \pi \sigma_{\parallel} & -D_{nE}k^2 \\ - 4 \pi \sigma'_\parallel & i \omega - D_{EE}k^2 \end{pmatrix} \begin{pmatrix} \delta n \\ \delta n_E \end{pmatrix} = 0.
\end{align}
Longitudinal modes of charge and energy exist when the determinant condition
\begin{equation}
\begin{vmatrix} i \omega - 4 \pi \sigma_{\parallel} & -D_{nE}k^2 \\ - 4 \pi \sigma'_\parallel & i \omega - D_{EE}k^2 \end{vmatrix} = 0.
\end{equation}
is satisfied. The roots of this equation yield dispersion relations
\begin{align}
\label{eq:longphot}
\omega = \begin{cases} -i\left[4\pi\sigma + \left(D_{nn} + \frac{\sigma'}{\sigma} D_{nE}\right)k^2 + \mathcal{O}(k^4) \right] \\
-i\left(D_{EE} - \frac{\sigma'}{\sigma}D_{nE}\right)k^2 + \mathcal{O}(k^4)
\end{cases}
\end{align}
Notice that the charge mode decays at a rate $4\pi \sigma$, and is not hydrodynamic as $\omega \to 0$; this is the rapid relaxation of free charge that is familiar in ordinary conductors. The energy mode is diffusive, just as in the paramagnetic phase. From Eq. \eqref{eq:longmodes}, it follows that the energy mode is accompanied by the charge fluctuation $\delta n = -\left(\frac{D_{nE}k^2}{4\pi \sigma}+\mathcal{O}(k^4)\right) \delta n_E$; by the constitutive relations \eqref{eq:jparr} and \eqref{eq:jeparr}, these give rise to fluctuations of longitudinal charge and energy current. Thus the diffusive energy mode is a simultaneous excitation of the modes $\{\delta n_E, \delta n, \delta \mathbf{e}_{\parallel}, \delta \mathbf{j}_{\parallel}, \delta \mathbf{j}_{E,\parallel}\}$.

\subsubsection{Summary}
We deduce that the $U(1)$ spin liquid with a Fermi surface has five diffusive hydrodynamic modes in total, given to leading order in $k$ by (restoring magnetization modes)
\begin{align}
i\omega_{S^{\alpha}} &= D_{SS} k^2, \quad \alpha=1,2,3, \\
i\omega_E &= \left(D_{EE} - \frac{\sigma'}{\sigma}D_{nE}\right)k^2, \\
i\omega_{\mathbf{a}} &= D_{\mathbf{a}}k^2,
\end{align}
with linear response transport coefficients defined as above. 

\subsection{Hydrodynamic modes of the $SU(4)$ algebraic spin liquid}
The key difference compared to the hydrodynamics of the $U(1)$ spin liquid with a Fermi surface, beyond the additional hydrodynamic conservation laws arising from flavour symmetry that were discussed in the main text, is the emergence of momentum as a hydrodynamic mode. In Fourier space, the linear-response momentum modes satisfy
\begin{align}
\label{eq:ptrans}
(i\omega - \nu k^2)\delta \mathbf{P}_{\perp} &= 0, \\
\label{eq:plong}
(i\omega - (\nu+\zeta) k^2)\delta \mathbf{P}_{\parallel} &= ik\left[\frac{\partial p}{\partial n} \delta n + \frac{\partial p}{\partial n_E} \delta n_E \right].
\end{align}

\subsubsection{Transverse modes}
The only change in the transverse mode structure compared to the spinon Fermi surface state is an additional transverse mode of momentum, whose dispersion is given by
\begin{equation}
i \omega_{\mathbf{P},\perp} = \nu k^2
\end{equation}
This mode corresponds to a constant pressure excitation of $\{\delta \mathbf{P}_{\perp}\}$ and is decoupled from the other hydrodynamic modes.

\subsubsection{Longitudinal modes}
The longitudinal slow modes of the algebraic spin liquid consist of charge, energy and longitudinal momentum, $\delta P_{\parallel} = \hat{\mathbf{k}} \cdot \delta \mathbf{P}_{\parallel}$. Assuming an ultrarelativistic equation of state $p=p(n_E)=n_E/2$ at half-filling, the linear-response mode coupling equations are given by
\begin{align}
\label{eq:longmod}
\begin{pmatrix} i \omega - 4 \pi \sigma_{\parallel} & -D_{nE}k^2 & 0 \\
- 4 \pi \sigma'_\parallel & i \omega - D_{EE}k^2 & -i{c^{*}}^2k \\
0 & -ik/2 & i(\omega - \omega_{\mathbf{P},\parallel}(k))
\end{pmatrix} \begin{pmatrix} \delta n \\ \delta n_E \\ \delta P_{\parallel}\end{pmatrix} = 0,
\end{align}
where $\omega_{\mathbf{P},\parallel} = (\nu+\zeta)k^2$. To understand the effect of the coupling to longitudinal momentum compared to Eq. \eqref{eq:longmodes}, it is clearest to first discard dissipative terms (i.e. pass to the ballistic scaling limit), which yields
\begin{equation}
\begin{pmatrix} i \omega & 0 & 0 \\
0 & i \omega & -i{c^{*}}^2k \\
0 & -ik/2 & i\omega 
\end{pmatrix} \begin{pmatrix} \delta n \\ \delta n_E \\ \delta P_{\parallel}\end{pmatrix} = 0.
\end{equation}
The resulting determinant condition reads
\begin{equation}
\omega \left(\omega^2 - {c^*}^2k^2/2\right) = 0,
\end{equation}
from which we deduce that at the Euler scale, fluctuations of energy and longitudinal momentum combine to form ballistic ``sound'' modes, with speed
\begin{equation}
c_s = \frac{c^*}{\sqrt{2}}.
\end{equation}
The full determinant condition resulting from Eq. \eqref{eq:longmod} is rather opaque, but diffusive broadening of the sound modes can be obtained perturbatively in $k$ and reads
\begin{equation}
\omega_s = \pm c_s k - \frac{i}{2}\left(D_{EE}-\frac{\sigma'}{\sigma}D_{nE}+\nu+\zeta\right)k^2 + \mathcal{O}(k^3).
\end{equation}

\subsection{The effect of a Chern-Simons term}
\label{sec:AppB}
Given our results for $U(1)$ spin liquid states with an emergent Maxwell field, a natural question is whether additional hydrodynamic modes due to gauge symmetry arise in spin liquid states with a Chern-Simons term for the emergent gauge field, for example the long-anticipated gapped chiral spin liquid phase\cite{WWZ,KalmeyerLaughlin,Baskaran,TriangleExpt}, which has lately come into numerical focus for Heisenberg models on Kagome and triangular lattices\cite{ChiralKagome,Sheng15,White,Sheng16,LauchliCSL,AaronPaper}. In contrast to the ``Maxwell'' spin liquids considered above, we find that the presence of a Chern-Simons term generically prevents the propagation of hydrodynamic modes of the emergent gauge field.

\subsubsection{Equations of motion for the gauge field}
For concreteness, we work in $2+1$D with metric signature $(-++)$ and let $A^\mu = (\phi,\mathbf{a})$ be a $U(1)$ gauge field minimally coupled to some matter field $\psi$. The Maxwell-Chern-Simons action
\begin{equation}
\label{eq:MCSaction}
S[A,\psi] = \int d^3 x \, \left(- \frac{c}{16 \pi} F_{\mu \nu}F^{\mu \nu} + \frac{\kappa}{2\pi}
\epsilon^{\mu \nu \rho} A_{\mu} \partial_{\nu} A_\rho\right) + S_{\mathrm{matter}}[A,\psi] \end{equation}
yields equations of motion
\begin{equation}
\partial_\mu F^{\mu \nu} + \frac{2 \kappa}{c} \epsilon^{\nu \mu \rho} \partial_\mu A_\rho = - \frac{4 \pi}{c} J^{\nu}
\end{equation}
for the gauge field $A$, where we write $J^{\mu} = (n c, \mathbf{j}) = \delta S_{\mathrm{matter}}/\delta A_{\mu}$. In components, this reads
\begin{align}
\label{eq:divE}
\nabla \cdot \mathbf{e} &= 4 \pi n - \frac{2\kappa}{c} b, \\
\label{eq:curlB}
\nabla \times \mathbf{b} &= \frac{1}{c} \partial_t \mathbf{e} + \frac{4 \pi}{c} \mathbf{j} - \frac{2\kappa}{c} \begin{pmatrix} e_2 \\ -e_1 \end{pmatrix},
\end{align}
while the constraint $\epsilon^{\mu \nu \rho}\partial_{\mu}\partial_{\nu}A_{\rho}=0$ implies Faraday's law
\begin{equation}
\label{eq:Faraday}
\partial_t b + c(\partial_1 e_2 - \partial_2 e_1) = 0.
\end{equation}

\subsubsection{Hydrodynamic modes}
As discussed in the main text, spin liquid states in condensed matter systems generically inherit four hydrodynamic modes from the high-temperature paramagnetic phase, namely energy and the three components of spin. The possibility of additional hydrodynamic modes from spatial translation symmetry and global $U(1)$ symmetry of the spin liquid effective action was also discussed above. 

Here, we address the question of whether the (bulk) $U(1)$ gauge symmetry of Eq. \eqref{eq:MCSaction} can yield an additional hydrodynamic mode. With no Chern-Simons term, $U(1)$ gauge symmetry gives rise to the hydrodynamic transverse photon mode that was discussed in the main text. When a Chern-Simons term is added, this hydrodynamic mode acquires a gap.

Let us verify this point directly. We find that the usual wave equation for $b$ is modified by a massive source term,
\begin{align}
\left(\nabla^2 - \frac{1}{c^2}\partial_t^2\right) b = - \frac{4\pi}{c} \nabla \times \mathbf{j} - \frac{8\pi \kappa n}{c} + \left(\frac{2 \kappa}{c}\right)^2 b,
\end{align}
while the wave equation for $\mathbf{e}$ is modified to
\begin{align}
\left(\nabla^2 - \frac{1}{c^2}\partial_t^2\right) \mathbf{e} = 4 \pi \left(\nabla n + \frac{1}{c^2}\partial_t \mathbf{j} \right) + \frac{8 \pi \kappa}{c^2} \begin{pmatrix} j_2 \\ -j_1 \end{pmatrix} + \left(\frac{2\kappa}{c}\right)^2 \mathbf{e},
\end{align}
and therefore also acquires a gap. At the level of the linear-response magnetohydrodynamic approximation discussed in the main text, it is clear that these mass terms for $b$ and $\mathbf{e}$ can only be cancelled by a fine-tuning of the transport coefficients appearing in $n$ and $\mathbf{j}$, and thus the hydrodynamic transverse photon mode is generically suppressed in the presence of a Chern-Simons term.

\end{document}